\documentclass[aip,rsi,reprint,amsmath,amssymb]{revtex4-1}
\usepackage[utf8]{inputenc}
\usepackage{nicefrac}
\usepackage{graphicx}% Include figure files
\usepackage[shortlabels]{enumitem}
\usepackage{placeins}
\usepackage[squaren]{SIunits}
\usepackage{lineno}
\usepackage{tabularx}
\usepackage[version=4]{mhchem}
\usepackage{hyperref}
    \hypersetup{colorlinks=true, linkcolor=blue, breaklinks=true,  urlcolor= blue, citecolor=blue}

\begin{document}

%\preprint{}

\title{Correction of the VIR-visible data set from the Dawn mission at Vesta} %Title of paper



\author{B. Rousseau}
\email[]{batiste.rousseau@inaf.it}
\affiliation{IAPS-INAF, via Fosso del Cavaliere 100, 00133, Rome, Italy}
\author{M. C. De Sanctis}
\affiliation{IAPS-INAF, via Fosso del Cavaliere 100, 00133, Rome, Italy}
\author{A. Raponi}
\affiliation{IAPS-INAF, via Fosso del Cavaliere 100, 00133, Rome, Italy}
\author{M. Ciarniello}
\affiliation{IAPS-INAF, via Fosso del Cavaliere 100, 00133, Rome, Italy}
\author{E. Ammannito}
\affiliation{Italian Space Agency (ASI), Via del Politecnico, 00133, Rome, Italy}
\author{P. Scarica}
\affiliation{IAPS-INAF, via Fosso del Cavaliere 100, 00133, Rome, Italy}
\author{S. Fonte}
\affiliation{IAPS-INAF, via Fosso del Cavaliere 100, 00133, Rome, Italy}
\author{\mbox{A. Frigeri}}
\affiliation{IAPS-INAF, via Fosso del Cavaliere 100, 00133, Rome, Italy}
\author{F. G. Carrozzo}
\affiliation{IAPS-INAF, via Fosso del Cavaliere 100, 00133, Rome, Italy}
\author{F. Tosi}
\affiliation{IAPS-INAF, via Fosso del Cavaliere 100, 00133, Rome, Italy}


\date{\today}

% Abstract
\begin{abstract}
% insert abstract here (250 words max)
\vspace{0.2cm}
The following article has been accepted by Review of Scientific Instruments on 15 November 2020. After it is published, it will be found at \href{https://publishing.aip.org/resources/librarians/products/journals/}{this link}. \href{https://doi.org/10.1063/5.0022902}{doi: 10.1063/5.0022902}\vspace{0.5cm}\par
This work describes the correction method applied to the dataset acquired at the asteroid (4) Vesta by the visible channel of the Visible and InfraRed mapping spectrometer (VIR). The rising detector temperature during data acquisitions in the visible wavelengths leads to a spectral slope increase over the whole spectral range. This limits the accuracy of the studies of the Vesta' surface in this wavelength range. Here we detail an empirical method to correct for the visible detector temperature dependency while taking into account the specificity of the Vesta dataset.
\end{abstract}
\maketitle %\maketitle must follow title, authors, abstract and \pacs
%
% Introduction
\section{Introduction}
\label{Introduction}
The NASA Dawn spacecraft orbited the asteroid (4) Vesta from July 2011 to September 2012 \cite{2007_Russell}. This allowed the on board Visible InfraRed mapping spectrometer (VIR)\cite{2011_De_Sanctis_a} to nearly map the whole surface of the asteroid.\par
% VIR instrument
The VIR spectrometer is made of two channels which may operate both in synergy or separately: the visible channel (VIS), working in the \unit{0.25-1.07}{\micro\meter} spectral interval by using a charge coupled device (CCD) and the IR channel (IR), covering the \unit{1.02-5.09}{\micro\meter} range with a HgCdTe array.\par
The data acquired by the visible channel of VIR (VIR-VIS) suffer from the increase of the CCD temperature ($T_{VIS}$) during the period of data acquisition. This study addresses their correction thanks to a method already developed for the second target of the Dawn mission, Ceres, and detailed in \citet{2019_Rousseau_c}. In Sect. \ref{VIR data} we present the VIR-VIS dataset acquired at Vesta. Sect. \ref{Instrumental_issue} describes the effect of the increase of the CCD temperature on VIR-VIS data. Sect. \ref{Correction} provides details about the method adopted to correct the data, in particular in some specific cases. Examples of the correction are given is Sect. \ref{Results}, before conclusions as summarized in Sect. \ref{Conclusion}.
\begin{figure}[b]
    \includegraphics{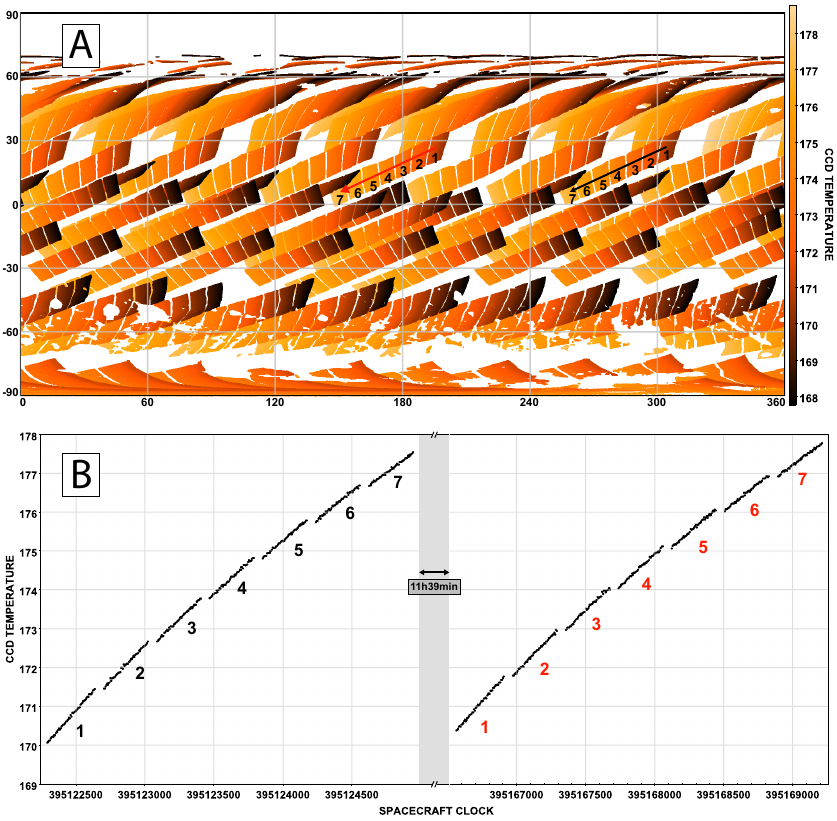}
    \caption{\label{FIG1_CCD_TEMP} Panel (A): Map of the CCD temperature for the VH2 mission phase. The different frames correspond to the projection of the field of view (FOV) of each hyperspectral cube. The increase of the CCD temperature is illustrated through two sequences of seven cubes each (black and red arrow). Panel (B): Evolution of the CCD temperature for the cubes of the two sequences indicated in Panel (A).}
\end{figure}
\begin{figure}[t]
    \includegraphics{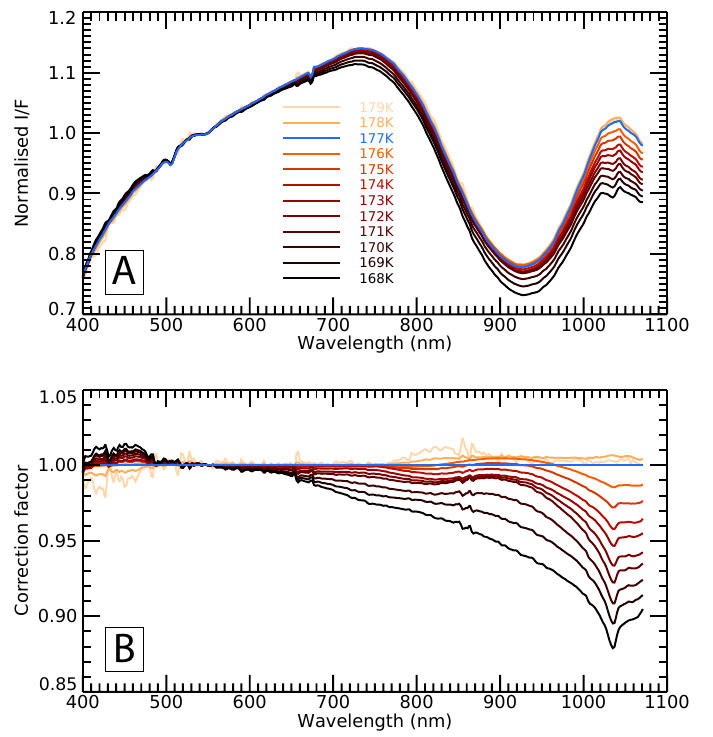}%
    \caption{\label{FIG2_spec_CF} Panel (A): Normalized median spectra from observations acquired during the VH2 mission phase for each interval of CCD temperature, illustrating the spectral reddening linked to the CCD temperature increase. The blue spectrum is the spectral reference as described in the text. Panel B: Correction factors relative to VH2 observations. The color scale is the same as that of Panel A. }%
\end{figure}
%
% Vesta Data
\section{The VIR data at Vesta}
\label{VIR data}
The Vesta dataset is divided in several mission phases whose the main characteristics are detailed in the Table \ref{Table1} (see also \citet{2011_Russell}).\par
The data acquired at Vesta by the VIR-VIS channel are available online on the Planetary Data System archive \footnote{\url{https://sbn.psi.edu/pds/resource/dawn/dwnvvirL1.html}}. We use the LEVEL 1B data which are calibrated from raw digital numbers (DN) to physical unit of radiance and then converted in radiance factor as detailed by \citet{2016_Carrozzo}.\par
We applied a correction for spectral artifacts \cite{2016_Carrozzo} and a photometric correction \cite{2016_Scarica} which minimizes the effect of the observation geometry. Finally, a correction factor developed by \citet{2016_Carrozzo} and refined by \citet{2020_Rousseau_a} is used to correct the shape and the systematic red slope observed in the VIR spectra of Vesta and Ceres. This latter correction factor, based on ground-based observations, is a multiplicative factor independent of the CCD temperature and, therefore, different from the one introduced in the present study.\par
We presently consider the spectral range from \unit{400}{\nano\meter} to its end.  However, we note that for spectral study of the surface, one would consider the shape of the spectra beyond \unit{1000}{\nano\meter} to be less reliable due to lack of signal and would therefore discard this wavelength range.
%
% TABLE
\begin{table*}[t]
\caption{\label{Table1} Mission phases of Dawn at Vesta, chronologically sorted and during which VIR-VISible data were acquired. Asterisks in the first column indicates that a correction factor has been computed for this mission phase (see last column). The fourth column reports the number of cubes processed and available (differences are due to the occurrence of sky observation or corrupted data). The resolution corresponds to the approximate minimum and maximum across-track projected pixel size on the surface. The $T_{VIS}$ and $T_{IR}$ columns provide the range of temperatures experienced by the visible and the infrared detectors respectively. The last column indicates the correction factor used to correct the mission phase.}
\resizebox{\textwidth}{!}{
\begin{tabularx}{17cm}{X X X X X X p{2.5cm} X}
\hline
\begin{tabular}{@{}l@{}}Mission\\ Phase\end{tabular} & \begin{tabular}{@{}l@{}}Start date\\yy-mm-dd\end{tabular} & \begin{tabular}{@{}l@{}}Stop date\\yy-mm-dd\end{tabular} & \begin{tabular}{@{}l@{}}Cubes\\(used/total)\end{tabular} & \begin{tabular}{@{}l@{}}Resolution\\ (m/pix)\end{tabular} & \begin{tabular}{@{}l@{}}$T_{VIS}$ (K)\end{tabular} & \begin{tabular}{@{}l@{}}$T_{IR}$ (K)\end{tabular} & \begin{tabular}{@{}l@{}}Correction\\factor\end{tabular} \\[5pt]
\hline \hline
VSA* & 2011-05-10 & 2011-08-06 & 96/133 & 700--1300 & 169--193 & 80 & VSA \\[2.5pt]
VSS* & 2011-08-12 & 2011-08-31 & 242/271 & 675--715 & 168--191 & 80 & VSS \\[2.5pt]
VTH & 2011-09-19 & 2011-09-25 & 8/12 & 170--205 & 171--180 & 80 & VH2 \\[2.5pt]
VSH* & 2011-09-30 & 2011-10-31 & 325/330 & 168--181 & 168--181 & 80 & VSH \\[2.5pt]
VSL* & 2012-01-08 & 2012-04-29 & 566/572 & 45--75 & 170--177 & 80 - 164 - 177 & VSL \\[2.5pt]
VH2* & 2012-06-15 & 2012-07-24 & 682/685 & 160--205 & 168--179 & 80 & VH2 \\[2.5pt]
VTC & 2012-08-25 & 2012-08-26 & 15/26 & 1485--1585 & 170--191 & 80 & VSS \\
\hline
\end{tabularx}
}
\end{table*}
%
% Instrumental issue
\section{Effects of the detector temperature variation}
\label{Instrumental_issue}
During an observation sequence, the VIR instrument may operate for some hours consecutively. In this temporal interval, several hyperspectral cubes are acquired and, in the case of the VIS channel, the temperature of the CCD progressively increases before passively cooling down when the acquisition sequence stops. The Fig. \ref{FIG1_CCD_TEMP} illustrates this phenomenon through a map of the CCD temperature during the Vesta High Altitude Mapping Orbit 2 (VH2) mission phase. We observed that the spectral response of the CCD is dependent on its temperature. One of the consequences is an apparent increase of the overall spectral slope of the VIR-VIS spectra, which we refer to as "reddening". This is illustrated by the Panel (A) of Fig. \ref{FIG2_spec_CF} which groups spectra acquired at the same CCD temperature during the VH2 mission phase.\par
In contrast, the IR channel is actively cooled down and its temperature ($T_{IR}$) is generally stabilized around 80K. However, if it is not operating, the cryocooler is switched off and the IR detector temperature gently increases until reaching an equilibrium temperature. It has been observed that small spectral distortions affect the VIR-VIS data acquired when $T_{IR}$ is high, i.e. not stabilized around 80K \citet{2019_Rousseau_c}. In the case of the Dawn mission at Vesta, this is limited to a portion of the VSL mission phase (see Table \ref{Table1}).\par
These issues limit the accuracy of the studies of the Vesta' surface in the visible range and consequently need to be corrected. Here we take advantage of the empirical correction developed to correct the VIR data acquired at Ceres, which suffers from the same issues \cite{2019_Rousseau_c}, and we adapt it to the Vesta dataset.\par
\begin{figure}[b]
    \includegraphics{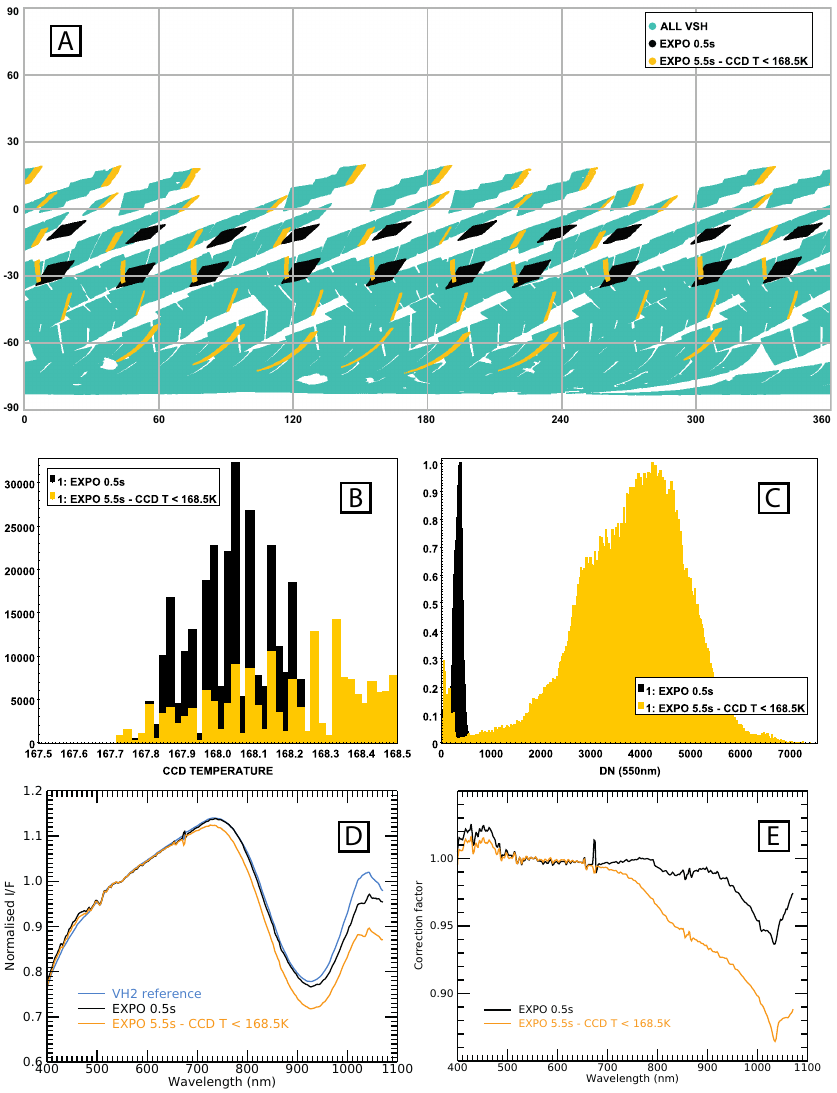}%
    \caption{\label{FIG3_MAP}Panel (A): Map of the VSH mission phase. The color of the footprints correspond to: in black, cube acquired with an exposition time of \unit{0.5}{\second} -- those cubes have a $T_{VIS}$ comprise in the 168K interval of CDD temperature (see text for details); in yellow, part of cubes acquired with an exposure time of \unit{5.5}{\second} and whose the $T_{VIS}$ is between 167.5K and 168.5 like the black one; in blue, cubes acquired with an exposure time of \unit{5.5}{\second} and whose $T_{VIS}$ is greater than 168.5K. Panel (B) to (E) represents respectively, for the yellow and black subsets (see Panel (A)): Panel (B): distribution of the $T_{VIS}$; Panel (C): distribution of the Digit Numbers at \unit{550}{\nano\meter}; Panel (D): corresponding median spectra -- the blue spectrum is the VH2 reference spectrum; Panel (E): corresponding correction factors.}%
\end{figure}
%
% Correction
\section{Correction of the data}
\label{Correction}
Using the same hypothesizes as for the Ceres case \cite{2019_Rousseau_c}, we developed a correction factor for different mission phases of the Vesta dataset. Those correction factors are based on a specific spectral reference, corresponding to a CDD temperature at which the detector response is considered reliable. We previously identified this CCD temperature to be equal to 177K\cite{2019_Rousseau_c} and the reference spectrum has been computed by taking advantage of the VH2 mission phase observations. In this respect, among the different mission phases, VH2 observations have the advantage \mbox{of a) a high redundancy} in the 176.5K and 177.5K CCD temperature interval ($799216$ spectra) and of b) being homogeneously distributed over the Vesta surface, providing a representative sample of its average spectral properties. Considering this, the generic formula of the correction factor (CF) is:
\begin{equation}
\label{eqn:CF}
CF_{X,\lambda,T_{VIS}}=\frac{(\nicefrac{I}{F})_{X,\lambda,T_{VIS}}}{(\nicefrac{I}{F})_{VH2, \lambda, T_{VIS}=177K}}
\end{equation}\par
Where $X$ corresponds to the name of the mission phase to be corrected; $\lambda$ is the wavelength; and $T_{VIS}$ the VIS CCD temperature. At the numerator, $I/F$ is the median radiance factor normalized at \unit{550}{\nano\meter} for the given mission phase to be corrected, wavelength, and detector temperature interval. At the denominator, $(I/F)_{VH2}$ is the median radiance factor, normalized at \unit{550}{\nano\meter}, of the observations acquired during VH2 at $T_{VIS}=177\text{K}$ and $T_{IR}=80\text{K}$. This is the blue reference spectrum as highlighted in Panel (A) of Fig. \ref{FIG2_spec_CF}.\par
The correction factor ($CF$) is calculated for discrete CDD temperature interval of 1K, and is a wavelength dependent vector. The set of correction factors of VH2 is represented in Panel (B) of Fig. \ref{FIG2_spec_CF}. For each spectrum, the correction factor to be applied is derived by interpolating $CF$ at the corresponding temperature of acquisition. It is applied by dividing this spectrum by the corresponding interpolated $CF$.\par
$CF$ for several mission phases have been specially calculated because:
\begin{enumerate}
    \item During the mission phases, the spatial sampling of the surface varies and implies slightly different median spectral behavior and, consequently, correction factors. This has to be considered to avoid an inappropriate $CF$ to be applied on a dataset.
    \item To avoid any correction errors, no extrapolation outside the minimum or the maximum CCD temperature bound is done (for a given mission phase), leading to the necessity of more than one $CF$ in case of CCD temperature ranges mismatch between different mission phases.
    \item We identify two mission phases (VSL and VSH) which must benefit from a dedicated correction factor (see Sects. \ref{VSH_Special_case} and \ref{VSL_Special_case}).
\end{enumerate} 
The Table \ref{Table1} reports the different mission phases with the corresponding correction factors to be applied. In the following sections, we provide details about the correction needed for the VSL and the VSH mission phases.
\begin{figure*}[t]
    \includegraphics[width=17cm]{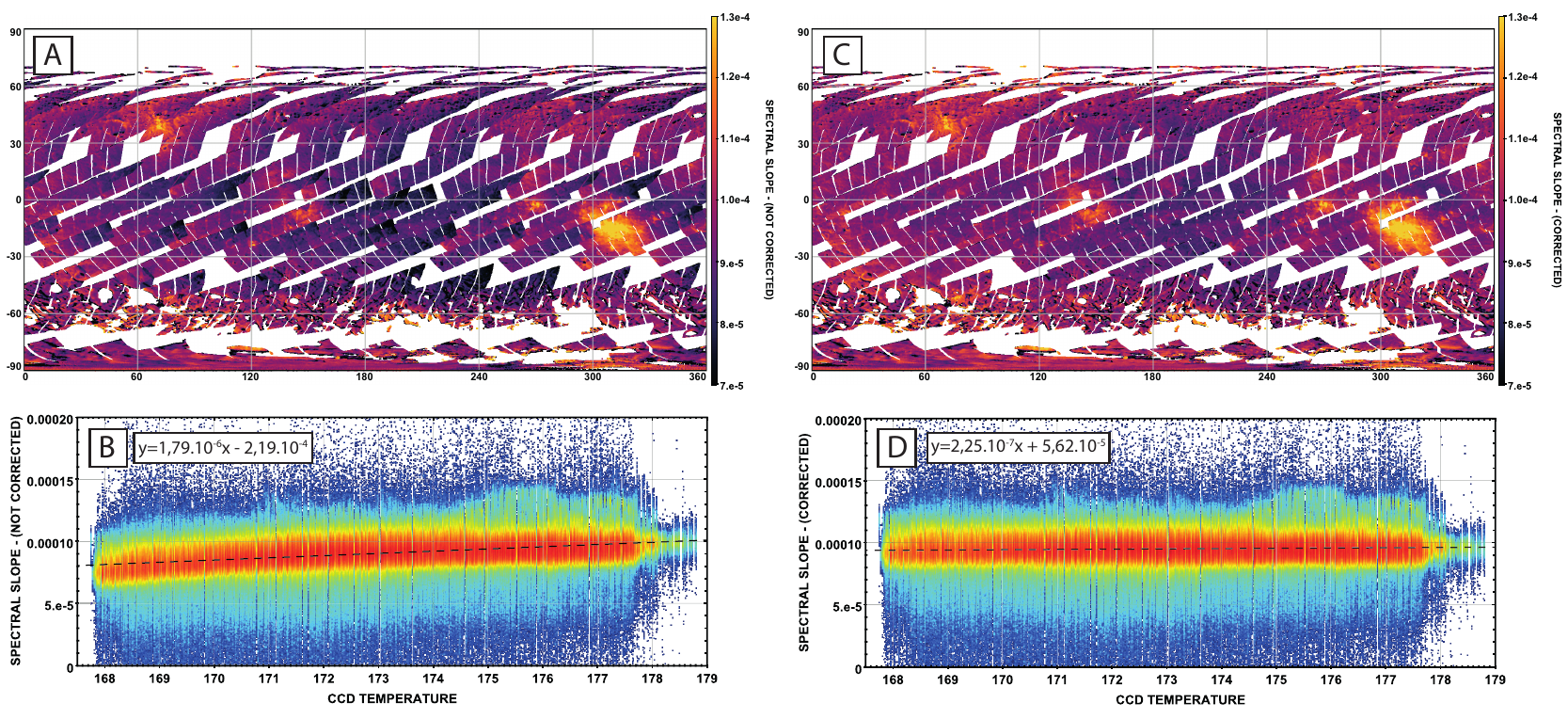}%
    \caption{\label{FIG4_PROOF_CORR}Panel (A): Map of the spectral slope $S_{465\text{--}730nm}$ before the correction. Panel (B): Spectral slope distribution with CCD temperature before correction. Panel (C): Map of the spectral slope $S_{465\text{--}730nm}$ after the correction. Panel (D): Spectral slope distribution with CCD temperature after the correction. See text for the $S_{465\text{--}730nm}$ definition. The equation of the linear fits for the Panel (B) and (D) is reported in their upper left corner.}%
\end{figure*}
%
% VSH and VSL specific cases 
%VSH
\subsection{Special cases: the VSH mission phase}
\label{VSH_Special_case}
The VSH mission phase covers mainly the south hemisphere of Vesta, as displayed by Panel (A) of Fig. \ref{FIG3_MAP}. The $T_{VIS}$ ranges from 168K to 181K and two exposure times, of \unit{0.9}{\second} and \unit{5.5}{\second}, have been used during VSH. Twenty cubes have been acquired with an exposure time of \unit{0.9}{\second} and are represented in black on Fig. \ref{FIG3_MAP}. Their $T_{VIS}$ is comprised in the 168K interval of CCD temperature (Panel (B) Fig. \ref{FIG1_CCD_TEMP}) as defined by the method to compute the correction factor, i.e. between 167.5K and 168.5K. What remains have been acquired with an exposure time of \unit{5.5}{\second} and spans various CCD temperatures. A part of these data has, just like the black subset, a $T_{VIS}$ in the 168K interval; displayed in yellow on Fig. \ref{FIG3_MAP}. As shows by the Panel (C) Fig. \ref{FIG3_MAP}, the two dataset, for the same $T_{VIS}$, have a level of signal very contrasted: the data with a low exposure time have a mean DN level at \unit{550}{\nano\meter} of 340 while the DN level of the second dataset, with a higher exposure time, spreads roughly between 1000 and 7000. This impacts the spectral behavior which are different between the two subsets (see the corresponding median spectra, Panel (D), Fig. \ref{FIG3_MAP}). This additional aspect must be taken take into account to avoid an incorrect correction as a single correction factor would have done. Consequently, a correction factor has been calculated for each of the subset identified in the 168K interval of CCD temperature (see Panel (E), Fig. \ref{FIG3_MAP}). 
%VSL
\subsection{Special cases: the VSL mission phase}
\label{VSL_Special_case}
About $21\%$ of the data in the VSL mission phase were acquired while the IR detector temperature ($T_{IR}$) was larger than 80K. A first group of observations was performed with $T_{IR}$ comprised between 163K and 165K and a second between 176K and 178K, as reported in Table \ref{Table1}. High IR temperatures induce a spectral distortion that has to be corrected too\cite{2019_Rousseau_c}. We adopted the same strategy as for VSH to correct those data, considering independently the two high IR temperature populations and computing separate correction factors: a first for the data acquired with a $T_{IR}$ close to 80K; a second for the data with a $T_{IR}$ around 164K; a third for the data with a $T_{IR}$ around 178K (see Supplementary Material). The use of the VH2 reference spectrum (see equation \ref{eqn:CF}), which is derived for an IR temperature of 80K, allows us to correct the spectral behavior of the VSL data with those high IR temperature.\par
\begin{figure}
    \includegraphics{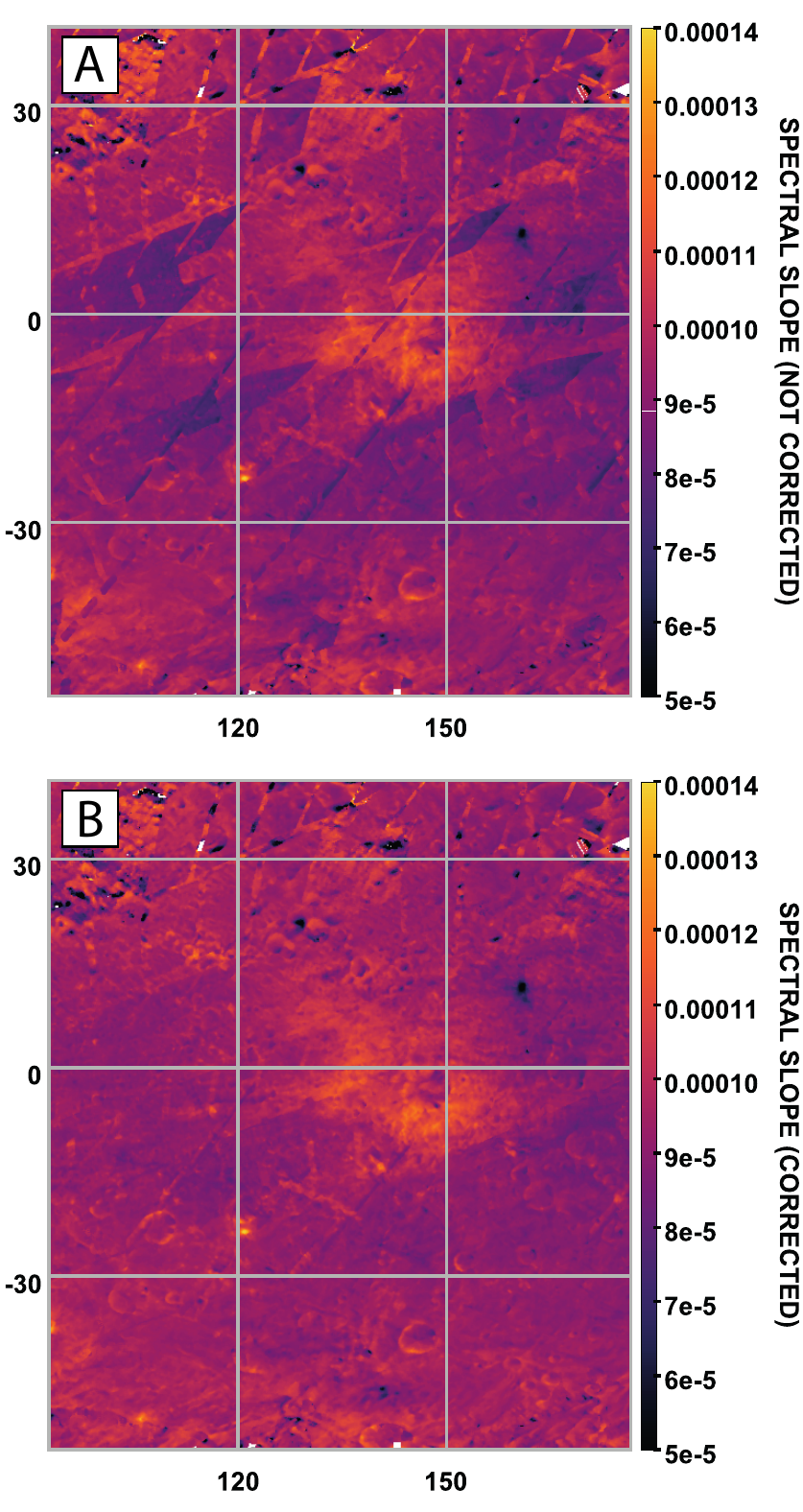}%
    \caption{\label{FIG5_ZOOM}Close-up of the spectral slope map before (Panel (A)) and after the correction (Panel (B)). The data which constitute the map come from all the mission phases.}%
\end{figure}
%
%Results
\section{Results}
\label{Results}
We estimate the efficiency of the correction through the spectral slope that we define as follow:
\begin{equation}
\label{eqn:SLOPE}
	S_{465\text{--}730nm}=\frac{(\nicefrac{I}{F})_{730nm}-(\nicefrac{I}{F})_{465nm}}{(\nicefrac{I}{F})_{465nm}\times(7300{\text{\normalfont\AA}}-4650{\text{\normalfont\AA}})}
\end{equation}
Where $S_{465\text{--}730nm}$ ($S$ hereafter) is the spectral slope computed between the radiance factor at \unit{730}{\nano\meter} ($(I/F)_{730nm}$) and at \unit{465}{\nano\meter} ($(I/F)_{465nm}$) and is expressed in \reciprocal{\kilo\angstrom}. This spectral indicator is particularly adapted to evaluate the correction since the evolution of the CCD temperature affects the global slope of the VIR-VIS spectra.\par
The Fig. \ref{FIG4_PROOF_CORR} reports the correction of the entire VH2 dataset. Maps of the spectral slope and spectral slope distributions with temperature, before and after the correction, are presented. The map of the Panel (A) corresponds to $S$ without correction. A gradient in the spectral slope, due to the increase of $T_{VIS}$ (see Fig. \ref{FIG1_CCD_TEMP}), is visible throughout a sequence of cubes. This gradient is also visible in the Panel (B) which present $S$ according to $T_{VIS}$. In this spectral slope distribution, the correlation coefficient of the linear fit is equal to $0.361$ while it drops to $0.048$ after the application of the correction (Panel (E)). This, combined with the lower slope of the linear fit after correction, indicates the accuracy of the latter which is also illustrated through the map of Panel (C).\par
The correction of the different mission phases of the VIR-VIS Vesta dataset allows their merging for mapping purposes. In the Fig. \ref{FIG5_ZOOM}, we present an example of such exercise through a close-up of the Vesta' surface. The maps are built with approximately 3.4M pixels and represent $S$ with data from all the mission phases. Pixels are projected as single point and the median is computed in case of overlapping. The Panel (A) shows the spectral slope before the correction; VIR image cube footprints from diverse mission phases are visible due to the absence of correction. On the contrary, Panel (B) exhibits a map almost devoid of artifacts thanks to the correction, even if some footprints are still slightly visible.\par
%
% Conclusion
\section{Conclusion}
\label{Conclusion}
The data acquired at Vesta and Ceres by the visible channel of the VIR spectrometer suffer from variations in the CCD temperature during their acquisition. The data acquired at Ceres have already benefited from an empirical correction detailed by \citet{2019_Rousseau_c}. Here, we developed a similar correction adapted for the Vesta dataset.\par
Such correction is necessary to carry out reliable studies of the surface like: mapping of spectral parameters (spectral slopes, reflectance), spectral analysis (especially using various VIR hyperspectral cubes), joint studies with the VIR infrared channel (merging of the data) or with the Dawn Framing Camera, photometric correction or analysis (the observation geometry is known to play a role on the spectral slope), spectral modeling for retrieving e.g. the composition.\par
The correction detailed in this present paper will be included in the calibrated version of the VIR visible data that will be delivered to the PDS (see the link in Sect. Data Availability Statement) in a future data release.
\section*{Supplementary Material}
See supplementary material for the figures of the median spectra and correction factors of the VSA, VSS, VSH, VSL and VTC mission phases.

\section*{Data Availability Statement}
The data that support the findings of this study are openly available on the Planetary Data Systeme archive at \url{https://sbn.psi.edu/pds/resource/dawn/dwnvvirL1.html}.

\begin{acknowledgments}
VIR is funded by the Italian Space Agency (ASI) and was developed under the leadership of INAF-Istituto di Astrofisica e Planetologia Spaziali, Rome, Italy (Grant ASI INAF I/004/12/0). The instrument was built by Selex-Galileo, Florence, Italy. The authors acknowledge the support of the Dawn Science, Instrument, and Operations Teams.
The authors made use of TOPCAT (Tools for OPerations on Catalogues And Tables\cite{2005_Taylor}) for a part of the data analysis and figure production. We thank the two reviewers for the suggestions which improved the paper.
\end{acknowledgments}
\bibliography{00_MT.bbl}
\end{document}

%\preprint{}

\title{Supplementary material of "Correction of the VIR-visible data set from the Dawn mission at Vesta"} %Title of paper



\author{B. Rousseau}
\email[]{batiste.rousseau@inaf.it}
\affiliation{IAPS-INAF, via Fosso del Cavaliere 100, 00133, Rome, Italy}
\author{M. C. De Sanctis}
\affiliation{IAPS-INAF, via Fosso del Cavaliere 100, 00133, Rome, Italy}
\author{A. Raponi}
\affiliation{IAPS-INAF, via Fosso del Cavaliere 100, 00133, Rome, Italy}
\author{M. Ciarniello}
\affiliation{IAPS-INAF, via Fosso del Cavaliere 100, 00133, Rome, Italy}
\author{E. Ammannito}
\affiliation{Italian Space Agency (ASI), Via del Politecnico, 00133, Rome, Italy}
\author{P. Scarica}
\affiliation{IAPS-INAF, via Fosso del Cavaliere 100, 00133, Rome, Italy}
\author{S. Fonte}
\affiliation{IAPS-INAF, via Fosso del Cavaliere 100, 00133, Rome, Italy}
\author{A. Frigeri}
\affiliation{IAPS-INAF, via Fosso del Cavaliere 100, 00133, Rome, Italy}
\author{F. G. Carrozzo}
\affiliation{IAPS-INAF, via Fosso del Cavaliere 100, 00133, Rome, Italy}
\author{F. Tosi}
\affiliation{IAPS-INAF, via Fosso del Cavaliere 100, 00133, Rome, Italy}


\date{\today}
\maketitle %\maketitle must follow title, authors, abstract and \pacs

\begin{figure*}
    \includegraphics{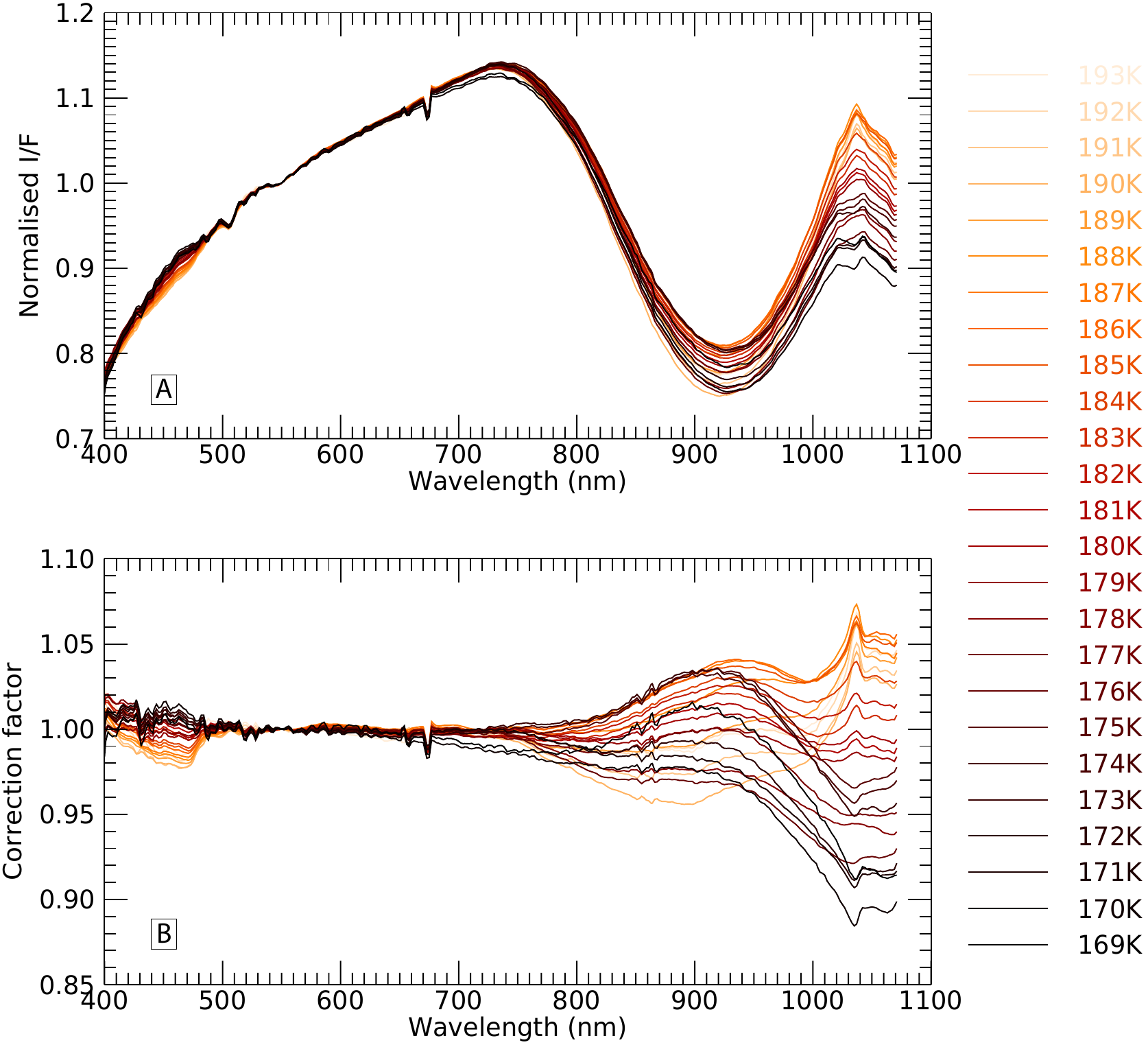}%
    \caption{\label{VSA}Panel (A): Median spectra, normalized at \unit{550}{\nano\meter}, for each bin of CCD temperature of the VSA mission phase. Panel (B): Corresponding correction factor. The CCD temperature color scale is the same for the two graphics.}%
\end{figure*}
\begin{figure*}
    \includegraphics{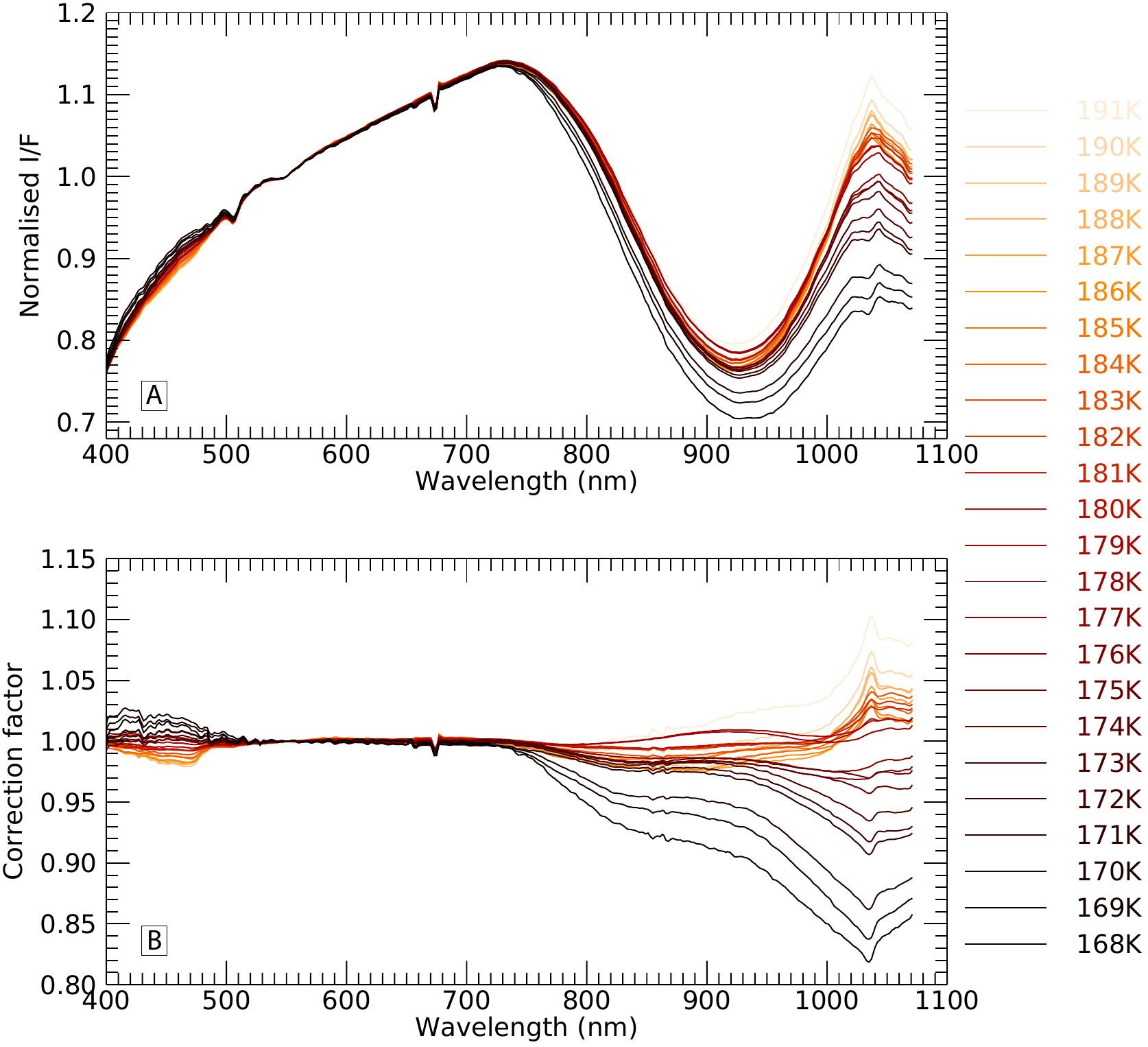}%
    \caption{\label{VSS}Same as Fig. \ref{VSA} for the VSS mission phase.}%
\end{figure*}
\begin{figure*}
    \includegraphics{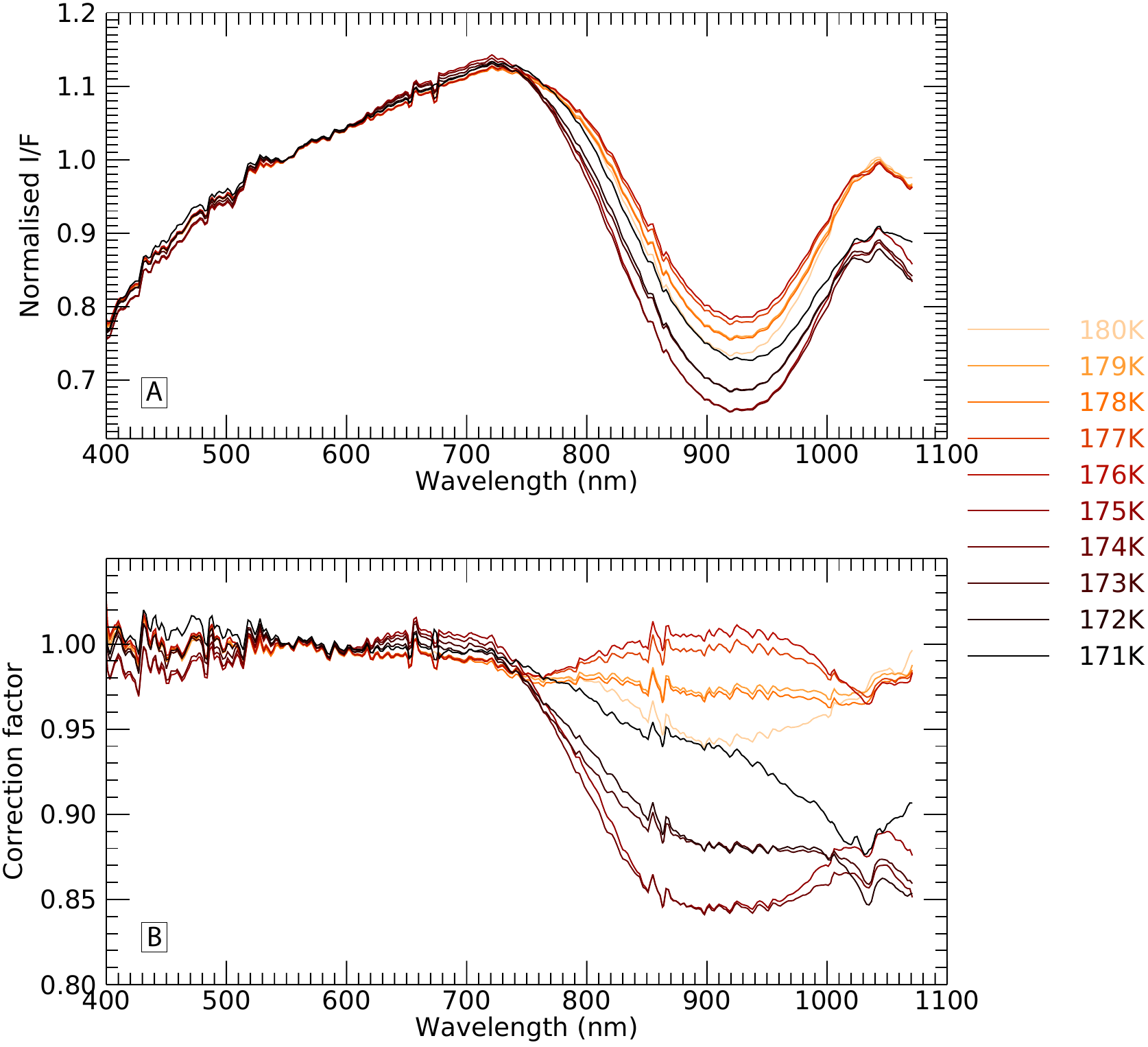}%
    \caption{\label{VTH}Same as Fig. \ref{VSA} for the VTH mission phase.}%
\end{figure*}
\begin{figure*}
    \includegraphics{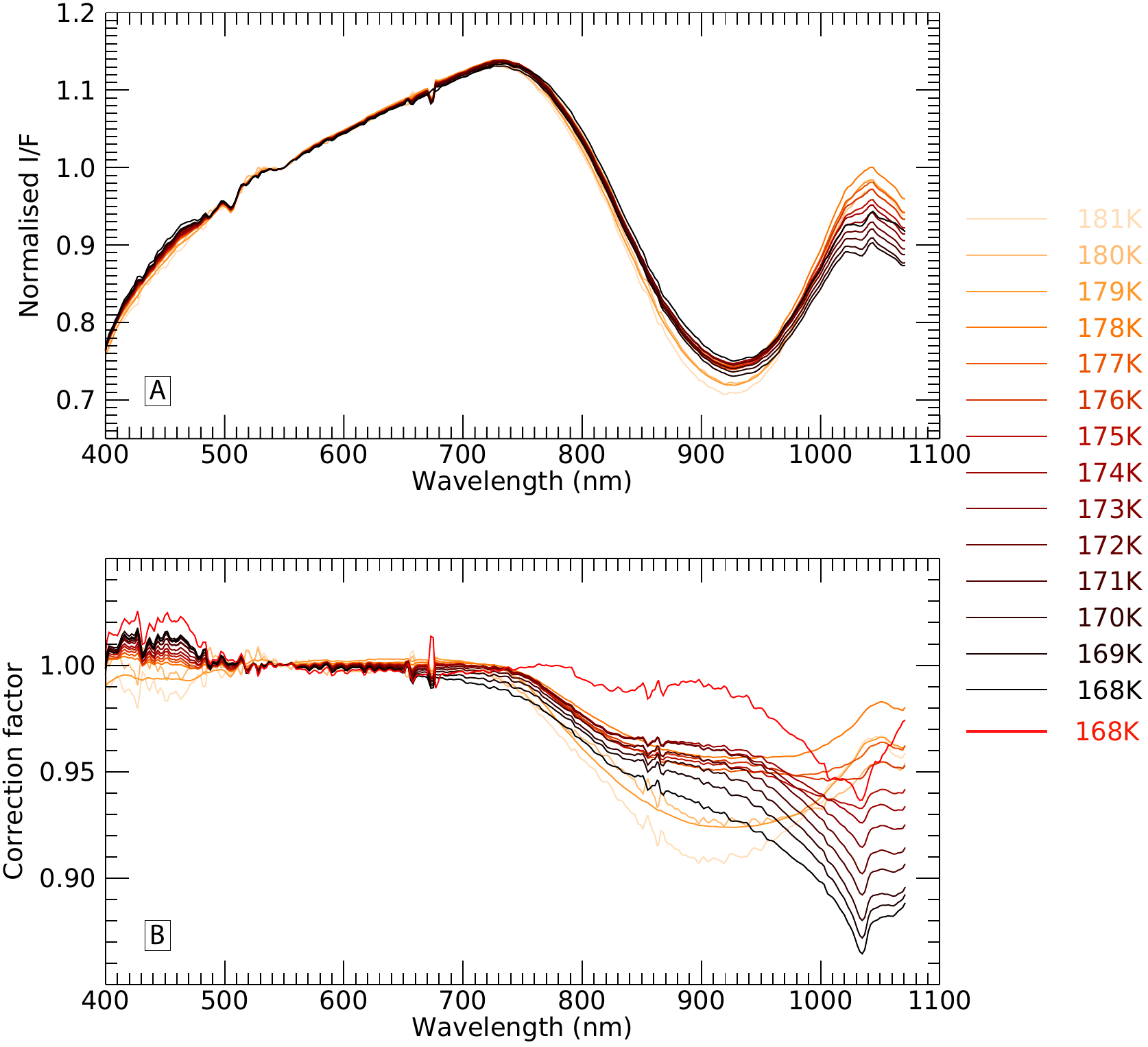}%
    \caption{\label{VSH}Same as Fig. \ref{VSA} for the VSH mission phase. The spectrum and the correction factor display in red correspond to the subset with low signal which need to be separated from the rest of the data to be properly corrected. See text for more details about the correction of the VSH mission phase.}%
\end{figure*}
\begin{figure*}
    \includegraphics{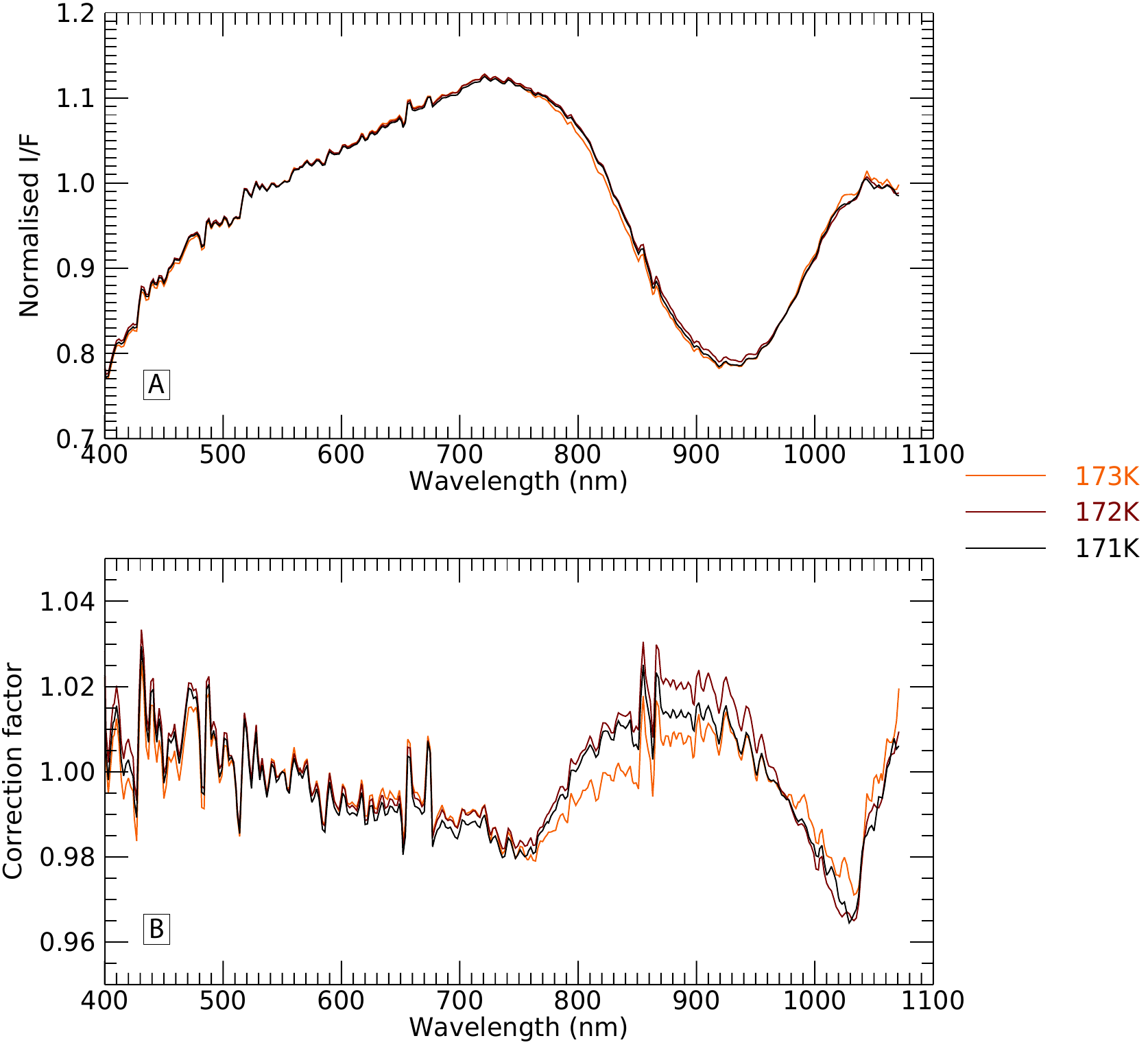}%
    \caption{\label{VSL_164}Same as Fig. \ref{VSA} for the VSL mission phase and for data acquired with $T_{IR}$ ($T_{IR}$ corresponds to the temperature of the infrared detector) around 164K. See text for more details about the correction of the VSL mission phase.}%
\end{figure*}
\begin{figure*}
    \includegraphics{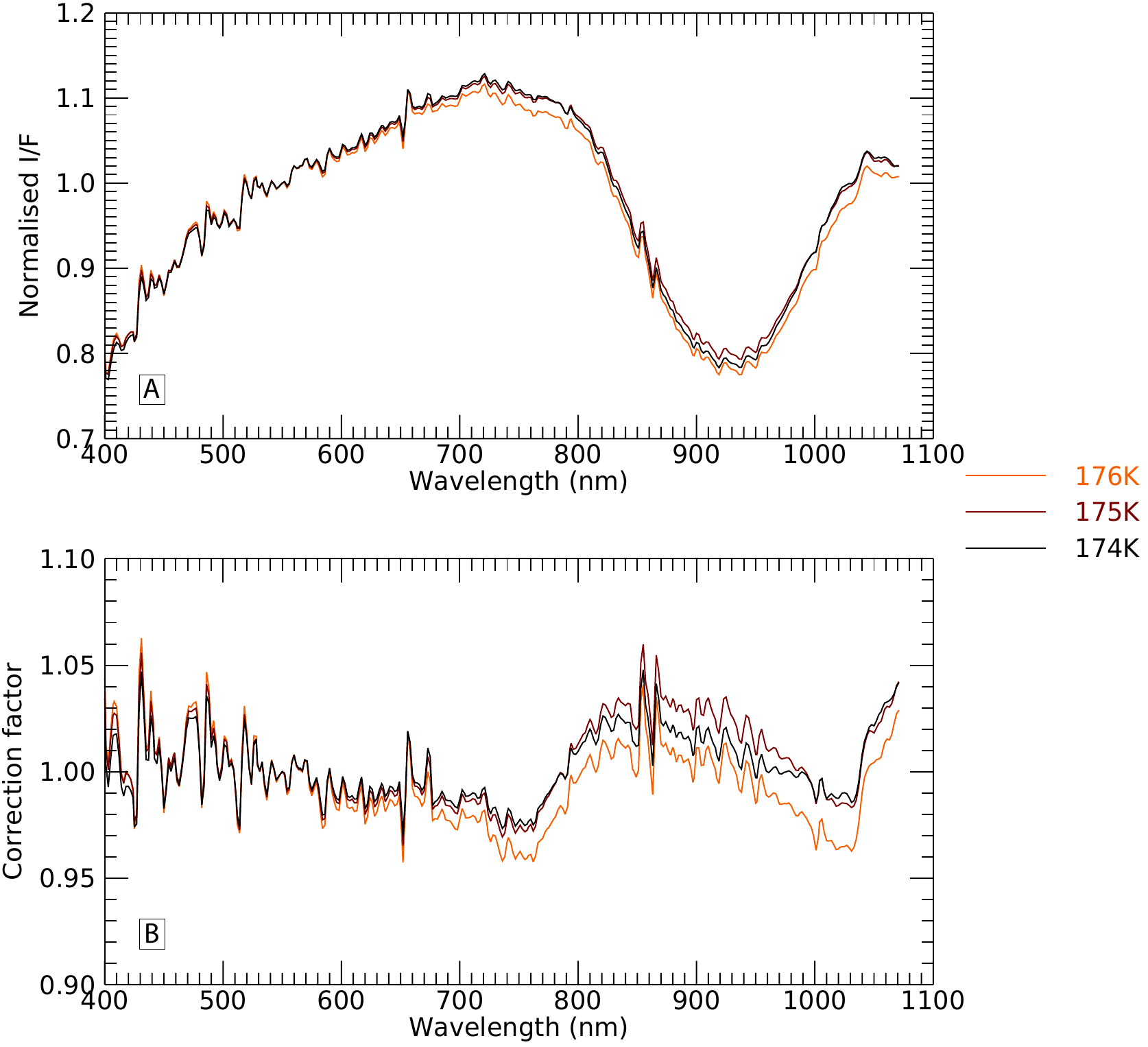}%
    \caption{\label{VSL_178}Same as Fig. \ref{VSA} for the VSL mission phase and for data acquired with $T_{IR}$ around 178K. See text for more details about the correction of the VSL mission phase.}%
\end{figure*}
\begin{figure*}
    \includegraphics{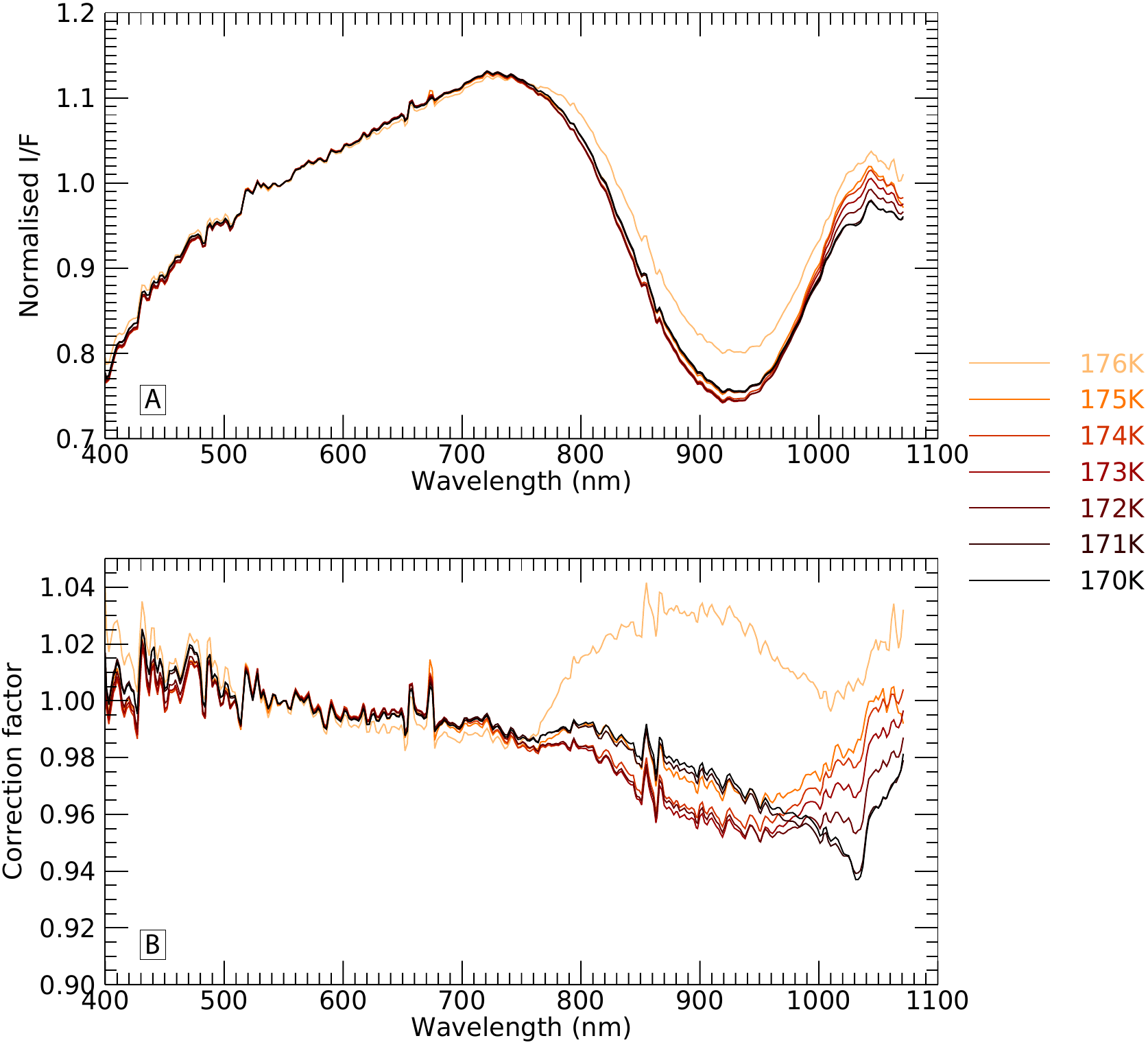}%
    \caption{\label{VSL_80_0}Same as Fig. \ref{VSA} for the VSL mission phase and for data acquired with $T_{IR}$ around 80K. See text for more details about the correction of the VSL mission phase.}%
\end{figure*}
\begin{figure*}
    \includegraphics{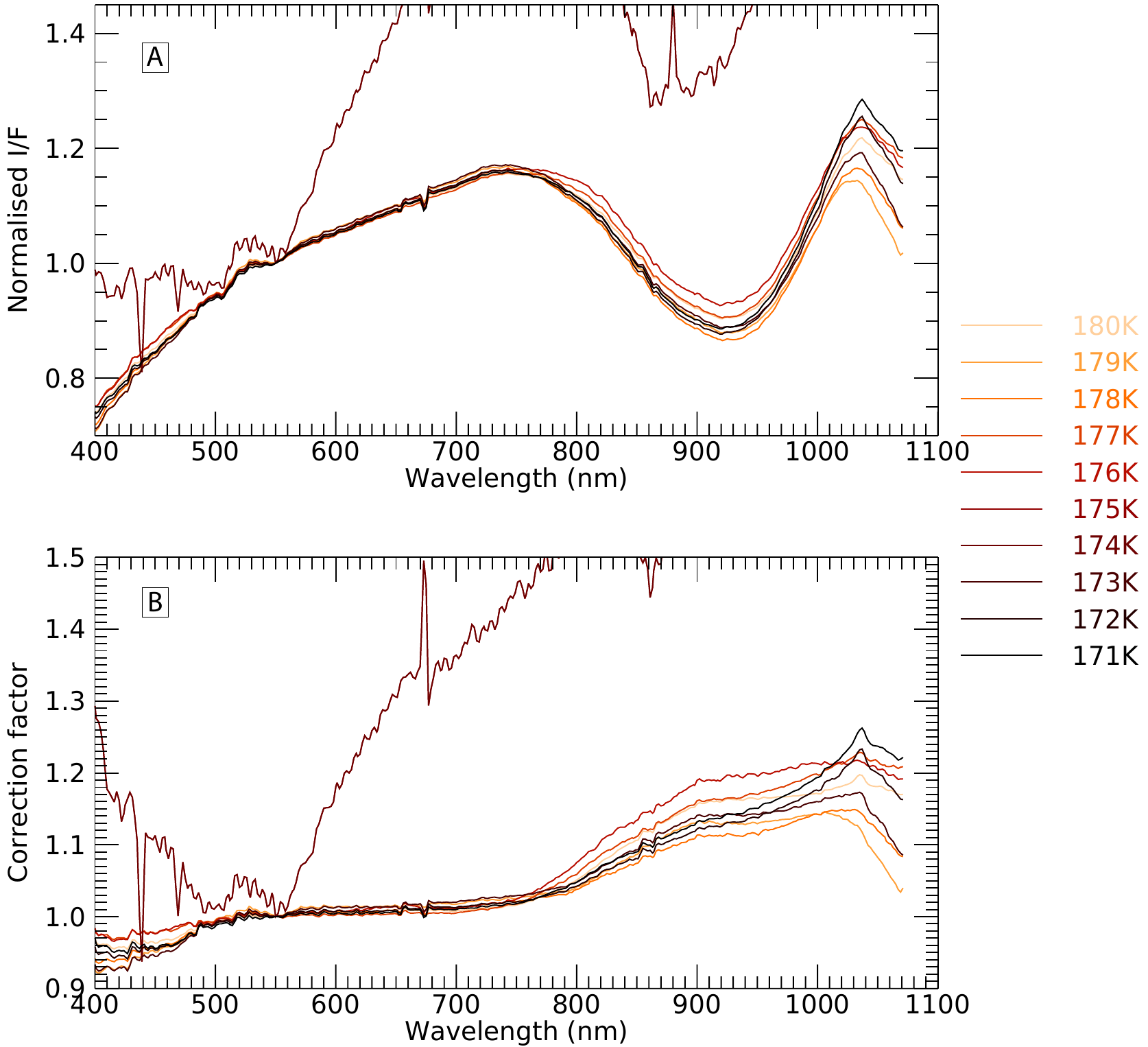}%
    \caption{\label{VTC}Same as Fig. \ref{VSA} for the VTC mission phase.}%
\end{figure*}